\documentclass[titlepage, 12pt]{article}

\usepackage{times}
\usepackage{url}
\usepackage{authblk}
\usepackage{setspace}
\usepackage{indentfirst}
\usepackage{natbib}
\usepackage[left=2.54cm, right=2.54cm, top=2.54cm, bottom=2.54cm
]{geometry} 
\usepackage{threeparttable}
\usepackage{booktabs}
\usepackage{multirow}
\usepackage[usestackEOL]{stackengine}
\usepackage{caption}
\usepackage{graphicx}
\usepackage [english]{babel}
\usepackage [autostyle, english = american]{csquotes}
\MakeOuterQuote{"}
\usepackage{amsmath}
\usepackage{tabularx}
\usepackage{titlesec}

\begin{document}
\title{The Impact of \textit{Pradhan Mantri Ujjwala Yojana} on Indian Households}

\author[1]{Nabeel Asharaf}
\author[1,2,3,4,5,6,7]{Richard S.J. Tol}
   
\affil[1]{Department of Economics, University of Sussex}
\affil[2]{Institute for Environmental Studies, Vrije Universiteit}
\affil[3]{Department of Spatial Economics, Vrije Universiteit}
\affil[4]{Tinbergen Institute}
\affil[5]{CESifo, Munich, Germany}
\affil[6]{Payne Institute for Public Policy, Colorado School of Mines, Golden, CO,USA}
\affil[7]{College of Business, Abu Dhabi University, UAE}

\maketitle

\begin{abstract}
This study critically evaluates the impact of the\textit{ Pradhan Mantri Ujjwala Yojana} (PMUY) on LPG accessibility among poor households in India. Using Propensity Score Matching and Difference-in-Differences estimators and the National Family Health Survey (NFHS) dataset, the Average Treatment Effect on the intendedly Treated is a modest 2.1 percentage point increase in LPG consumption due to PMUY, with a parallel decrease in firewood consumption. Regional analysis reveals differential impacts, with significant progress in the North, West, and South but less pronounced effects in the East and North East. The study also underscores variance across social groups, with Scheduled Caste households showing the most substantial benefits, while Scheduled Tribes households are hardly affected. Despite the PMUY’s initial success in facilitating LPG access, sustaining its usage remains challenging. Policy should emphasise targeted interventions, income support, and address regional and community-specific disparities for the sustained usage of LPG.
\end{abstract}

\newpage
\section{Introduction}
\par Access to clean energy is essential for a high-quality life, yet a significant portion of the global population faces challenges in obtaining clean energy for fundamental needs like lighting, heating, and cooking. While progress has been made in achieving the United Nations Sustainable Development Goal 7 of universal access to clean and affordable energy by 2030, persistent inequalities still exist \citep{undp2015}. Modern cooking fuels, crucial for health and environmental benefits, remain inaccessible for millions, particularly in developing regions. This discrepancy in progress between access to electricity and clean cooking fuels underscores the need for targeted policy interventions to address challenges hindering the adoption of modern cooking fuels. In this paper, we are the first to study one such policy intervention in India, the \textit{Pradhan Mantri Ujjwala Yojana}. We find a causal effect on LPG use that is positive and highly significant but small.

\par The global pandemic and recent geopolitical events have exacerbated the situation, leading to a surge in energy prices worldwide, mainly for LPG and crude oil \citep{IEAWEO2022}. The increase in the LPG price has impacted many developing countries severely. The energy price rise is predicted to force up to 100 million people to revert to using traditional fuels for cooking without effective interventions \citep{IEAWEO2022}. The Government of India, for example, reduced LPG subsidies at the start of the pandemic when global energy prices fell. However, when energy prices surged again, the government could not fully reinstate the subsidies. This added financial pressure on households, as they had to pay more for LPG cylinders \citep{sharma2021target}. Energy poverty has garnered significant attention in development literature over the past two decades (see below). This heightened focus can be attributed to the profound repercussions associated with the absence of accessible, affordable, and modern energy resources for essential needs such as lighting, heating, and cooking.

\par India has nearly completed the universal electrification of households. The electricity coverage in India stands at 97 percent \citep{agrawal2020state}. But around 660 million people lack access to clean cooking fuels \citep{indiaIEA}. This distinction is mainly because access to clean cooking encompasses more than just technical availability; it also encompasses considerations of dependability, ease of use, safety, and cost-effectiveness \citep{gould2018lpg}. Many policies were introduced to reduce the dependence on traditional cooking stoves in India, but most policies failed in achieving their goal \citep{khandelwal2017have}. Several factors contribute to the limited demand for modern cooking technologies. These include high initial cost and subpar stove quality, inadequately planned and executed policies, a disconnect between stove features and local requirements, and a lack of consideration for environmental and health issues by the general public \citep{mobarak2012low}. This is set against the backdrop of the statistics that approximately 17.8 \% of India's total fatalities in 2019 were attributed to indoor air pollution \citep{pandey2021health}.  That is, around 1.67 million deaths in India in the year 2019 were due to indoor air pollution. Encouraging households to transition away from traditional cooking stoves has consistently posed a significant policy challenge for successive Indian governments. The cost of LPG has been a contentious issue in political discussions, and the fiscal constraints associated with it have often made it difficult for the government to policy reform measures. 

\par
\begin{figure}[ht]
\caption{The Evolution Price/ Kg of Commercial and Household LPG}
\label{commercial}
\centering
\includegraphics[width=1\textwidth]{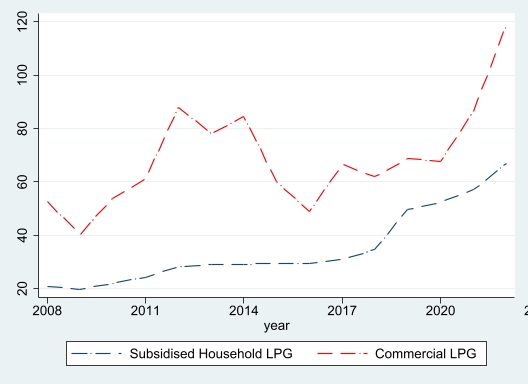}
\caption*{ \raggedright Source: \cite{lpgstat}, \cite{IOClstat}
\par \textbf{Note:} The Rs/Kg of 14.2 subsidised Household LPG and 19 Kg Commercial purpose LPG. The price represents the cost of an LPG cylinder in Delhi at the beginning of April each year}
\end{figure}

 LPG was provided to Indian households at a subsidised price from 1970 onwards. After the year 2000, when the Indian government initiated the deregulation of the LPG industry, there was a notable increase in LPG adoption among households. However, this also led to an upsurge in the sale of subsidised LPG gas cylinders to commercial buyers in the informal market \citep{agarwal2021evolution}. This period was also accompanied by increased fiscal pressure on the government due to the increased subsidy cost. The significant price disparity between subsidised household LPG and unsubsidised commercial LPG is one of the main reasons for household LPG being sold to commercial users in the informal market \citep{mittal2017fuel}. This price disparity between the 14.2 Kg subsidised household LPG and the 19 Kg commercial purpose LPG is shown in figure \ref{commercial}.

\par
\begin{figure}[ht]
\caption{Price Evolution of Subsidised and Non-Subsidised LPG Cylinders}
\label{subsidised}
\centering
\includegraphics[width=1\textwidth]{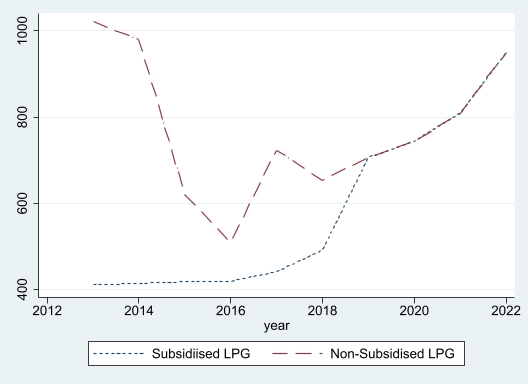}
\caption*{ \raggedright Source: \cite{lpgstat}, \cite{IOClstat}
\par \textbf{Note:} The cylinder's price is indicated in INR (Indian Rupees). The price represents the cost of an LPG cylinder in Delhi at the beginning of April each year}
\end{figure}

 To address the shortcomings of earlier policies and to increase the usage of LPG, various policies were introduced by the Government of India since 2012. These initiatives include programs such as PAHAL, which involve selling LPG at market prices while directly transferring subsidies to the recipient's bank account \citep{pena2019lpg}. Also, the number of cylinders consumers can buy a year is capped. The beneficiaries are now determined based on the household's income level and the universal subsidy for LPG consumption is scrapped. But from 2019, the direct bank subsidy transfer to the beneficiaries was completely phased out. This is graphically represented in the figure \ref{subsidised}. Despite the drop in crude prices at the onset of the Covid-19 pandemic leading to the discontinuation of the subsidy, it has not been restored even as prices have rebounded. As of August 2023, the price of an LPG cylinder in Delhi is INR 1103, which is roughly equal to USD 13.

\par Another policy to address the problem of low consumption of LPG among poor households was the\textit{Pradhan Mantri Ujjwala Yojana }(PMUY henceforth)\footnote{PMUY can be loosely translated as Prime Minister's Cooking Gas Scheme}, launched in 2016. According to \cite{pmuyabout}, the main goal of PMUY is to provide clean cooking fuel like LPG to rural and underprivileged households that previously relied on traditional cooking fuels such as firewood, coal, and cow-dung cakes. Under PMUY, a deposit-free LPG connection is provided to the adult women of the household. Under this scheme, the government covers the cost of various items, including the security deposit for the cylinder, pressure regulator, safety hose, DGCC booklet, and installation charges, amounting to a maximum of Rs 1600 per connection \citep{pib}. As of the Ministry's Official Report, there are 95 million PMUY beneficiaries, of which  84 million households have refilled at least once in the 2021-22 period \citep{pib1}. The Government has also approved a targeted subsidy of INR 200 to all PMUY beneficiaries, which is credited directly to the beneficiary's bank account starting from March 2023 \citep{pib2}.In this research study, we aim to evaluate the success of the PMUY in increasing LPG consumption among poor households in India.

\par The next section provides a brief review of the existing literature on energy poverty and the assessment of programs aimed at mitigating energy poverty. Subsequently, we delve into the discussion of data and the empirical methodology. The findings and their implications are discussed in the conclusion.

\section{Literature Review}
\par Energy poverty in India is widely discussed. \cite{balachandra2011} presents a comprehensive examination of the complexities surrounding energy poverty in rural areas of the country. By adopting a multidimensional approach, this study analyses the temporal, income, and regional aspects of energy access. The study brought attention to an imbalance in addressing energy access issues, where the focus on supply-side factors tends to outweigh the equally crucial consideration of demand-side aspects in academic research and policy formulation. Moreover, their descriptive analysis of NSSO data reveals limited progress in addressing energy poverty, especially among rural households, as negligible improvement can only be witnessed after 23 years of implementing the first policy to address it. According to the author, this is evident from low access levels among rural households and especially in households in the lowest income strata and significant variations in modern energy access levels across states in India. The main reasons for the lack of access among poor rural households are poor policy formulation and implementation, lack of resources and infrastructure and high poverty levels across the country.

\par  \cite{alkon2016household} conduct a detailed analysis using NSSO data from 1987 to 2010, investigating the variation of energy cost burdens among Indian households and the relationship between household expenditure and access to modern energy sources. The findings shed light on the energy cost burdens faced by different states, revealing that those with modernised power sectors experience lower energy cost burdens. In contrast, rural households encounter increased energy cost burdens despite improved energy access, as their income growth does not parallel urban counterparts. The study identifies access to electricity and LPG as influential factors positively associated with energy expenditure, highlighting those other factors beyond household income, such as government policies and market dynamics, play a role in influencing the adoption of modern fuel sources. Similarly, \cite{chindarkar2021examining} delves into the willingness of rural Indian households to pay for LPG, employing a probit model and leveraging data from the ACCESS 2018 dataset. The study assesses the probability of households exclusively using LPG for cooking for a specific bid amount for an LPG cylinder. The results indicate that as the randomly assigned bid for the LPG cylinder increases, the probability of households opting for exclusive LPG use decreases for both LPG and non-LPG households. Moreover, the study also showcased that salaried employees are more likely to accept bid amounts for exclusive LPG use than other households, citing the influence of regular income in the energy transition. 
  
\par Various studies have analysed the impact of the PMUY on LPG consumption in India such as \cite{ashwini,swain2020determinants,gill2020lpg,ranjan2020household}. Most of these studies are descriptive in nature or regional analysis. \cite{ranjan2020household} analysed the impact of PMUY using data from the National Sample Survey Office (NSSO) expenditure data and the Petroleum Planning and Analysis Cell (PPAC) report. The analysis reveals a notable decrease in firewood consumption in rural areas since 2014. Furthermore, the study's findings suggest that, based on the PPAC report, there has been a 15 \% increase in the consumption of LPG. However, they also found that the number of inactive LPG connections has risen. According to the study, this phenomenon can be attributed to PMUY's success in expanding access to LPG but its relative failure to encourage consistent and sustained usage among its beneficiaries. \cite{ashwini}, in their analysis, arrives at a similar conclusion, drawing upon information from the Petroleum Planning and Analysis Cell (PPAC) reports. They attribute the limited uptake in LPG refilling, even among those who gained access to LPG cylinders through PMUY, to the increased prices of LPG, including subsidised LPG cylinders.

\par In contrast, \cite{gill2020lpg}, utilising the ACCESS data from the Council on Energy, Environment, and Water (CEEW), examined the effects of the PMUY on LPG connectivity in six states. They employed an intent to treat Difference-in-Differences (DiD) methodology and arrived at the conclusion that PMUY had led to an increase in LPG connectivity, with gains ranging from 3.3 to 3.6 percentage points. Their control group consisted of households that slightly exceeded the income limit for the Below Poverty Line (BPL) classification. They also found that PMUY beneficiaries are far behind in terms of LPG consumption compared to the control group. \cite{swain2020determinants} conducted an analysis in the Indian state of Odisha to examine the factors influencing the adoption of LPG. They utilised both probit and tobit models for their analysis. Their key finding indicates that the reduction in firewood usage is more pronounced among households with higher levels of education but less significant among households benefiting from the PMUY. This outcome implies that PMUY may not have been entirely successful in expediting the energy transition in Odisha. It's worth noting that their primary survey data, collected from the Puri district in Odisha, comprises a relatively small sample of just 106 observations.

\par In conclusion, while the prevailing literature effectively underscores the importance of policies like the Pradhan Mantri Ujjwala Yojana (PMUY) within the Indian context, it is essential to acknowledge the predominant descriptive nature of most evaluations, often reliant on reports from the Petroleum Planning and Analysis Cell (PPAC). This reliance on descriptive assessments may hinder the establishment of a robust causal link to PMUY's impact. Additionally, assessments utilising the ACCESS dataset by CEEW are geographically limited to six Indian states with below-average per capita GDP figures, potentially limiting the external validity of the findings to other regions within the country. To address this gap in the literature, we focus on analysing the causal impact of the PMUY on LPG consumption in India. 

\section{Data and Descriptive Statistics}
\par The research relies on data from two waves of the National Family Household Survey (NFHS) dataset: NFHS-4 (2014-15) and NFHS-5 (2019-20). NFHS is a comprehensive survey conducted nationwide in India to gather crucial information on population, health, and nutrition. The main goal of NFHS surveys, as stated in their official document, is to provide essential data on health and family welfare, including crucial indicators like fertility levels, infant and child mortality, maternal and child health, and other emerging health and family welfare issues at both the national and state levels \nocite{iips2021nfhs} (IIPS and ICF, 2021). These surveys cover around 707 districts in all 29 states and 7 union territories of India and survey around 600,000 households in each wave. We here exclude the union territories.

\subsection{Key variables}
\par The variables used for predicting treatment and matching households into treatment and control groups  are given in table \ref{tab:table 1}. These include standard household socio-economic characteristics that are almost unique to Indian households, such as caste religion, along with other characteristics.   

\begin{table}[h]
\centering
\captionsetup{font=bf} 
\caption{Key Variables}
\label{tab:table 1}
\begin{tabular}{@{}ll@{}}
\toprule
\multicolumn{1}{l}{\textbf{Variables}} & \multicolumn{1}{l}{\textbf{Description}}                                 \\
\midrule
hhid               & Household Identification Number                                           \\
State              & State where the household resides
                   \\
Age                & Age of the household head                                                 \\
Religion           & Religion of the household head                                            \\
Caste              & Caste of the household head                                               \\
BPL card            & A dummy variable indicating whether the household holds a BPL card (1) or not (0) \\
Wealth Index          & Wealth index relative to the state                                        \\
Educ           & Highest educational attainment of the household head                      \\
urban       & A categorical variable indicating the household is in urban (1) or rural area (2)          \\
Gender             & Gender of the household head. (1 for male and 2 for female)                                              \\
hhsize             & Total number of people living in the households                                  \\ \bottomrule

\end{tabular}
\end{table}

\par The treatment group is all households with a Below Poverty Line (BPL) card because PMUY is essentially targeted at BPL households. The BPL card is provided to every household who lives under INR 1000 in cities or INR 816 in rural areas in India \citep{gaur2020poverty}. From table \ref{BPLtable}, It is clear that the percentage of BPL households with LPG access has increased by more than two-fold. It was 22 percent in 2014-15 and increased to 45.13 percent in 2019-20. It is a compounded annual growth rate of 15.38 percent. At the same time, the LPG access of non-BPL households increased by only 6.9 percent CAGR annually. So, the main objective of this research is to analyse whether this increase in the growth rate is due to the PMUY. 

\begin{table}
\begin{threeparttable}
\captionsetup{font=bf} 
\caption{Proportion of Household having Access to LPG}
\label{BPLtable}
\begin{tabularx}{\textwidth}{XXX}
\toprule
\textbf{Category} & \textbf{2014-15} & \textbf{2019-20} \\
\hline
BPL Household & 22.07 & 45.13 \\
Non-BPL Household & 44.32 & 61.88 \\
\hline
 \end{tabularx}
    \small{Source: Based on author's calculation using NFHS-4 and NFHS-5}
    \end{threeparttable}
\end{table}
\par
To simplify the analysis, India's 28 states are grouped into 6 distinct zones, each with its own unique characteristics. These zones are as follows:

\begin{enumerate}
  \item \textbf{North Zone:}  Haryana, Himachal Pradesh, Jammu and Kashmir, Punjab, and Rajasthan.
  \item \textbf{North East Zone:} Assam, Arunachal Pradesh, Manipur, Meghalaya, Mizoram, Nagaland, Tripura, and Sikkim.
  \item \textbf{Central Zone:} Chattisgarh, Madhya Pradesh, Uttarakhand, and Uttar Pradesh.
  \item \textbf{East Zone:}  Bihar, Jharkhand, Orissa, and West Bengal.
  \item \textbf{West Zone:}  Goa, Gujarat, and Maharashtra.
  \item \textbf{South Zone:}  Andhra Pradesh, Karnataka, Kerala, Tamil Nadu, and Telangana.
\end{enumerate}

These zones are characterised by their geographical, cultural, and socio-economic differences. In the regressions, we use \textit{state} fixed effects. The \textit{zones} are used for sample splits.  Table \ref{zonelpg} indicates the proportion of households having access to LPG in each zone. While all zones have an incremental increase in LPG access, the West Zone shows a decrease in the proportion of households having access to LPG for cooking. The Central Zone has the highest number of households with LPG access, followed by the South Zone.
\begin{table}[h]
\begin{threeparttable}
\captionsetup{font=bf}
\caption{Proportion of Households having access to LPG by Zone}
\label{zonelpg}
\begin{tabularx}{\textwidth}{XXX}
\toprule
\textbf{Zone} & \textbf{2014-15} & \textbf{2019-20} \\
\hline
North Zone & 40.88 & 61.51 \\
North East Zone & 32.18 & 44.75 \\
Central Zone & 30.43 & 69.91 \\
East Zone & 17.77 & 58.97 \\
West Zone & 46.74 & 33.32 \\
South Zone & 56.23 & 63.04 \\
\hline
\end{tabularx}
    \small{Source: Based on author's calculation using NFHS-4 and NFHS-5}
    \end{threeparttable}
\end{table}

\par \textit{Wealth Index} is a variable encompassing the household's assets and their composition. The index is determined by assigning scores to households based on the quantity and types of consumer products they possess. These products can include items such as televisions and automobiles, as well as household amenities like access to clean drinking water and sanitation facilities, as well as the materials used for flooring \nocite{iips2021nfhs} (IIPS and ICF, 2021). The index scores are then estimated using principal component analysis by the DHS. It has five categories: poorest, poorer, middle, richer and richest. More than 50 percent of the rural population in India falls into the lowest two categories: the poorest and the poorer. For the analysis, the wealth index within the state is used as it is more meaningful to compare the poor households within the state rather than comparing with the poor in other states.
 
\begin{table}
\begin{threeparttable}
\captionsetup{font=bf}
\caption{Summary Statistics}
\label{summ}
\begin{tabularx}{\linewidth}{lXXXXXX}
\hline
\multicolumn{7}{c}{Summary Statistics} \\
\multicolumn{1}{c}{\multirow{2}{*}{Variables}} & \multicolumn{3}{c}{Treatment group} & \multicolumn{3}{c}{Control Group} \\
\cmidrule(lr){2-4} \cmidrule(lr){5-7} 
\multicolumn{1}{c}{} & Obs & Mean & S.D & Obs & Mean & S.D \\
\hline
Age & 459,940 & 49.370 & 13.770 & 564,532 & 48.576 & 14.485 \\
HHsize & 459,940 & 4.593 & 2.208 & 564,532 & 4.524 & 2.223 \\
Urban & 459,940 & 1.817 & 0.387 & 564,532 & 1.644 & 0.479 \\
Gender & 459,940 & 1.177 & 0.382 & 564,532 & 1.150 & 0.357 \\
Cooking fuel & 459,940 & 6.247 & 4.989 & 564,532 & 5.314 & 6.040 \\
Wealth Index & 459,940 & 2.446 & 1.279 & 564,532 & 3.158 & 1.434 \\
Religion & 459,940 & 3.141 & 12.816 & 564,532 & 2.648 & 10.450 \\
Caste & 444,953 & 2.452 & 1.052 & 542,232 & 2.782 & 1.123 \\
LPG Access & 459,940 & 0.337 & 0.473 & 564,532 & 0.506 & 0.500 \\
\hline
\end{tabularx}
\small
Note: Household belongs to the BPL category is the treatment group and not belong to the BPL category is the control group. Summary statistics is estimated before the matching exercise.
\end{threeparttable}
\end{table}

\par Table \ref{summ} shows the summary statistics of the model. The treatment group is the households below the poverty line, and the control group is the households above. The total number of observations for the treatment group is around 450,000 and around 550,000 for the control group, which is the non-BPL category. The mean difference between the treatment and control group is not much different for most of the variables except for the cooking fuel.

\section{ Empirical Strategy}
\par Due to a lack of information on PMUY beneficiaries in the NFHS dataset, an Intention to Treat (ITT) analysis is utilised, following the approach outlined by  \cite{duflo2007using}. The ITT approach considers all eligible individuals as part of the treatment group, irrespective of whether they received the treatment. The households with a Below Poverty Line (BPL) card are designated as the treatment group, while those without are the control group. Propensity scores, derived from observable characteristics, group households into treatment and control groups. As the NFHS dataset is no panel, the analysis adopts a repeated cross-sectional approach. A propensity Score Matching Method is employed to estimate the causal impact of PMUY. 

\subsection{Propensity Score Matching}
\par The Propensity Score Matching (PSM) method is the most commonly used methodology to replicate the randomisation process in policy evaluation exercises where randomised data is not available. This approach pairs individuals from the control and treatment groups with similar propensity scores, effectively matching them based on these characteristics \citep{athey2017state}. The resulting treatment and control groups, formed through propensity score matching, enable a direct and balanced comparison (\citep{rosenbaum1983central}). The PSM method addresses potential bias and facilitates the creation of more balanced comparison groups. This single score reflects the likelihood of receiving treatment, allowing to mitigate bias by matching treated and untreated subjects more effectively \citep{rubin1997estimating}. 

\par The model relies on two essential assumptions: Conditional Mean Assumption (CIA) and overlap. CIA assumes that observed outcome differences between treatment and control groups arise solely from the treatment, effectively controlling for other influences. Overlap, on the other hand, ensures that individuals in both groups have a non-zero probability of receiving either treatment or control, enabling meaningful comparisons. These assumptions are essential in assuming that the treatment satisfies (some form) of exogeneity. The observations are matched into treatment and control groups based on the predicted propensity score. The nearest neighbour matching algorithm pairs units in the treatment and control groups. Further, the quality of treatment exercise (balancing property) is also examined. To examine the balancing property of the matching exercise, \cite{rubin2001using} suggested three measures: analysing the standardised bias and examining Rubin's B and  Rubin's R. The results are provided in table \ref{tab:bias}.
 
\par The matching is carried out in both pre and post-treatment periods. A Difference in Differences (DiD) estimator similar to \cite{heckman1997matching, heckman1998characterizing} is employed to estimate the treatment effect on the treated. So, the ATT is estimated for the pre-treatment period using PSM. Then, the ATT is estimated for the post-treatment period. Then, the DiD is employed to calculate the difference between the ATTs. In this way, the bias arising out of time-invariant unobservables can be isolated \citep{smith2005does}.

\par The treatment assignment equation employed in the model is
\begin{equation}
     prob[Treatment=1]= \frac{exp(z_i)}{1+exp(z_i)}
\end{equation}
where:
\begin{multline}
\label{treatment eqn}
  z_i = \gamma_0+ \gamma_1state_i + \gamma_2hhsize_i + \gamma_3urban\_rural_i + \gamma_4age_i + \gamma_5religion_i + \gamma_6caste_i + \\ \gamma_7bplcard_i + \gamma_8wi\_state_i + \gamma_9hheduc_i + \gamma_{10}gender_i
\end{multline}
where subscript \textit{i} represents the household. The treatment is 1 if the household belongs to the BPL category and 0 if it does not belong to the BPL category. All other explanatory variables are explained earlier in the data section. 

There are two outcome variables in this model. 
\begin{enumerate}
     \item LPG Access: 1 if the household has LPG, 0 otherwise
     \item Firewood Access: 1 if the household uses firewood for cooking, 0 otherwise
\end{enumerate}
\par Despite the advantage of propensity score matching for emulating randomisation in observational studies, it is essential to recognise that PSM has its own limitations. A significant concern is due to the reduction in dimensionality. Combining various covariates into a single propensity score, although helpful for predicting treatment assignment, can potentially reduce the overall effectiveness of the analysis. Another critical consideration revolves around PSM's exclusive reliance on observable confounding covariates. It is vital to acknowledge that PSM does not accommodate unobservable confounding covariates, a potential source of bias that could impact the accuracy and validity of the estimated outcomes \citep{rubin1997estimating}.

\subsection{Robustness Analysis}
\par In order to check the robustness of the results computed using the PSM methodology, Inverse Probability Weight and Augmented Inverse Probability Weight (AIPW) Estimators are used to compute the causal impact of PMUY using the same dataset. Even though PSM reduces bias by adjusting the propensity score for treatment and control groups, it decreases the dimensions of the estimate because reducing covariates into a single propensity score is less efficient \citep{hirano2003efficient}. To overcome this flaw, IPW estimators give each observation a weight, which is the inverse of the probability of treatment. The idea is that the observation within the treatment group having the highest weight is more similar to the control group because they have less propensity for treatment and therefore more inverse probability weight \citep{huntington2021effect}. That is, an observation which gets treatment will have a weight of $1/p$, whereas an observation in the control group will have a weight of $1/(1-p)$. The higher weight will be given to the treatment group observation, which has a low propensity to get treated (high inverse probability) but gets treated anyways and to control group observations, which have a high propensity to be in the treatment group but did not get the treatment (thus high inverse probability). This weighting exercise creates a pseudo-population that, once re-weighted, will demonstrate a comparable distribution of observable traits.

\par The augmented Inverse Probability Weighting (AIPW) estimator is a doubly robust estimator, which is a consistent estimator to estimate ATE when the treatment model is correctly specified or the outcome model is correctly specified \citep{glynn2010introduction}. It is called a doubly robust estimator because it estimates a consistent ATE if any of the two regression models, one with matched covariates and treatment as the dependent variable or the other regression model with the outcome as the dependent variable, are correctly specified \citep{huntington2021effect}. \cite{robins1994estimation}, first proposed this estimator based on inverse probability weights, which are asymptomatically normal and unbiased to estimate the unknown parameter in the regression equation when there is a problem of missing data in the model. The ATE of  AIPW is expressed in the equation \ref{equation 11}. 
\begin{multline} \label{equation 11}
     ATE_{aipw} = \frac{1}{n}\sum_{i=1}^{n} E(Y_i|D=1) - E(Y_i|D=0) + \\ \frac{(D_i)(Y_i - E(Y_i|D=1))}{e_i(x)} - \frac{(1-D_i)(Y_i-E(Y_i|D=0)}{1-e_i(x)}
\end{multline}
\par The AIPW estimator has many statistical properties similar to other estimators. As mentioned earlier, it is consistent if either the outcome model or the model to predict the propensity score is correctly specified. Because the bias in the first part of equation \ref{equation 11} will be estimated and eliminated by the second part of the equation \ref{equation 11} or vice versa. The other properties of AIPW include attaining variance bounds if both models are correctly specified, efficient if the outcome model is correctly specified and joint population boundness \citep{tan2010bounded}. The treatment assignment equation is the same as equation-\ref{treatment eqn}.
The same DiD estimator is employed to calculate the difference between the pre and post-treatment ATT to filter out the potential unobservable bias in the treatment effects for both IPW and AIPW estimators. 

\par Further, the sensitivity of treatment effect due to unobserved selection bias is also tested in this analysis using the \cite{rosenbaum2002overt} bound test. This test does not test for the magnitude of bias but instead tests how sensitive the treatment effect is to unobservable factors affecting the treatment assignment. Suppose the probability of treatment assignment is given by 
\begin{equation}
e(x) = prob(D_1 = 1|x,u_i) = F(\beta x_i + \gamma u_i)
\end{equation}
where $x_i$ is the observable characteristics, $u_i$ is the unobservable characteristics and $\gamma$ is the effect of the unobservables on the probability of the treatment. Suppose two units having the same probability of treatment, i.e.; $e(x_i) = e(x_j)$, the odds that two individuals getting treatment is given by  $ \frac{p_i}{1-p_i}$ and $\frac{p_j}{1-p_j}$. The odds ratio is then given by 
\begin{equation}
 \frac{\frac{{p_i}}{{1-p_i}}}{ {\frac{p_j}{1-p_j}}} = \frac{exp(\beta x_i)exp(\gamma u_i)}{exp(\beta x_j)exp(\gamma u_j)}
\end{equation}
Since the probability of getting treatment for the two units remains the same, the identical and observable $x$ vector gets cancelled out and the remaining part is $exp[(\gamma)( u_i- u_j)]$. If there is no unobservable selection bias or the unobservable has no impact on treatment assignment, the probability will be the same and the odds ratio will be 1. However, if the unobservable has an impact on the treatment probability or the unobservable affects the units differently, the odds ratio will not be equal to 1. \cite{rosenbaum2005sensitivity} suggest that in this case, the odds ratio will be bounded between
\begin{equation}
    \frac{1}{\Gamma} \leq \frac{p_i(1-p_j)}{p_j(1-p_i)} \leq \Gamma
\end{equation}
where $\Gamma = e^\gamma $ and assuming the unobservable factors affecting are binary in nature. 
\par The testing of sensitivity towards the unobservable factors is assigning different values for $\Gamma$ to examine for what values of $\Gamma$ the p-value of the odds of participation changes (probability ratio). If $\Gamma = 1$, then there are no unobservable factors affecting the odds of treatment ratio. And if $\Gamma = 2$, it means that the probability of treatment assignment is changed by a factor of two. That is, out of the two units, one unit's probability of getting treatment is doubled due to the unobservable confounders. In effect, the Rosenbaum Bound test manipulates the value of $\Gamma$ to gauge the extent to which hidden variables affect the odds of treatment assignment and subsequently, the study's conclusions. So if for a small value of $\Gamma$ ( say 1.5) the p-value changes, then the model is less robust \citep{imbens2009recent}. For binary outcome, the MH-stat similar to \citep{aakvik2001bounding} is used.

\section{Results and Discussions}
\subsection{Balancing Property Diagnostics}
\par Before discussing the treatment effects, it is crucial to analyse the balancing property of the matching assignment. For the matching at the baseline, i.e., for 2014-15, all the variables used for the treatment equation are highly significant except for the gender variable. That may be because only a small proportion of households out of the total households are headed by women. So, the gender variable might not be a good predictor of treatment for the model \footnote{The results of probit regression for the treatment assignment are provided in Appendix A}. Out of 570214 observations for the baseline, only 8 observations were off-support; the rest of the observations were on-support. Since the dataset is large and has 8 variables, kernel-based matching is beyond our computational power. So, the matching was based on the nearest neighbour algorithm without any caliper and replacement. 
\par To assess the balancing property of the matching, the standardised mean bias of each variable after matching is analysed. The mean bias indicates how well the distributions of covariates are balanced among the treatment and control groups. After matching, the mean of covariates should be similar for both the treatment and the control group for a fair comparison. Ideally, the standardised bias after matching should lie between -5 and 5. It is evident from table \ref{tab:bias} that all variables used to predict the treatment lie under the required bound. Another measure to check for good balancing is Rubin's R, which is the ratio of the variance of the propensity score for the treatment and control group. The recommended value of Rubin's R should lie between 0.5 and 2. And for all the variables in the treatment model, it lies within the bounds.   
\begin{table} [h]
\begin{threeparttable}
\caption{Balancing Property Diagnostics}
\label{tab:bias}
\begin{tabularx}{\textwidth}{XXXXX}
\toprule
\textbf{Variable} & \textbf{Treated} & \textbf{Control} & \textbf{\%bias} & \textbf{Rubin's R} \\
\midrule
State             & 17.855           & 18.021           & -2.1 &   1.06         \\
Age               & 48.906           & 48.614           & 2.3  &  0.94         \\
Religion          & 2.870          & 2.873            & -0.3    &  1.05        \\
Caste             & 2.475           & 2.482           & -0.6 & 1.03           \\
Educ          & 1.021           & 1.023          & -0.1 & 1.12          \\
Wealth Index         & 2.417           & 2.393           & 2.2  & 0.96         \\
urban\_rural      & 1.8134          & 1.825          & -2.6 & 1.04          \\
Gender            & 1.162           & 1.154           & 2.8 &   1.04        \\
hhsize            & 4.859           & 4.873           & -1.2 &   0.95         \\
\bottomrule
\end{tabularx}
\begin{tablenotes}[para,flushleft]
\small
\item Note: \%bias represents the percentage difference in means between treated and control groups. The ratio Rubin's R represents the variance ratio of the treated group to the control group.
\end{tablenotes}
\end{threeparttable}
\end{table}
\par 
\par
\begin{figure}[ht]
\caption{Probability Density of Treatment and Control Group}
\label{fig:mygraph}
\centering
\includegraphics[width=0.7\textwidth]{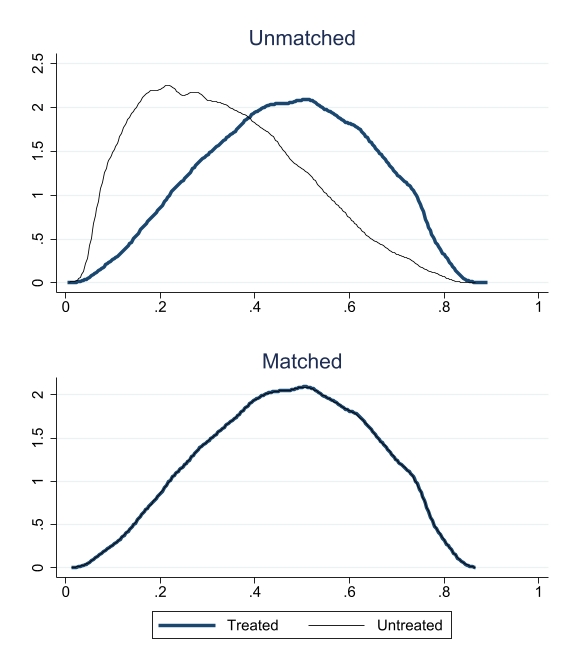}
\end{figure}

 Figure \ref{fig:mygraph} shows the density of propensity score before and after matching for treated and untreated groups. It is evident from the graph before matching that the density of propensity to get treated is different for both groups. However, once the observations are matched, the density is almost the same for both groups.  

\begin{table}[h]
\centering
\begin{threeparttable}
\captionsetup{font=bf}
\caption{Impact of PMUY on LPG Access}
\label{tab:result}
\begin{tabularx}{\textwidth}{Xllll}
\toprule
& \multicolumn{2}{c}{Full Sample} & \multicolumn{2}{c}{Poorest Sample} \\
\cmidrule(r){2-3} \cmidrule(r){4-5}
& Pre Treatment & Post Treatment & Pre Treatment & Post Treatment \\
\midrule
ATT (Treatment Group) & 0.220          & 0.451            & 0.026            & 0.194 \\
ATT (Control group)  & 0.230          & 0.439            & 0.035            & 0.165 \\
Difference in ATT & -0.010         & 0.012            & -0.009           & 0.029 \\
Standard Error & 0.002          & 0.003            & 0.002            & 0.004 \\
t-stat & -4.69          & 4.5              & -4.91            & 6.82 \\
\midrule
Treatment Effect & \multicolumn{2}{l}{0.021 ***	
} & \multicolumn{2}{l}{0.038***	
} \\
Standard Error & \multicolumn{2}{l}{0.003} & \multicolumn{2}{l}{0.005} \\
\midrule
On support Untreated & 349318       & 194726         & 62463           & 36898  \\
On support treated & 218753       & 226180         & 67979           & 68824 \\
\midrule
Mean Bias  & 1.6            & 2.1              & 1.2              & 1.2 \\
Rubin's B & 5.4            & 6.6              & 4.5              & 5.7 \\
Rubin's R  & 1.01           & 0.9              & 1.08             & 1.07  \\
\bottomrule
\end{tabularx}
\begin{tablenotes}[para,flushleft]
\item \textbf{Note}: Treatment effect is the difference in the ATT from the post and pre-treatment period. T-stat for the treatment effect is calculated using computing sample variance and standard error. The significance level is determined based on the t-stat. \item \textbf{***} $p < 0.01$, \textbf{**} $p < 0.05$,  \textbf{*} $p < 0.1$.
\end{tablenotes}
\end{threeparttable}
\end{table}

\subsection{ATT Estimates of PMUY}
\par Table \ref{tab:result} reports the ATT from the PSM matching exercise. The results suggest a causal impact of a 2.1 percentage point increase in the uptake of LPG among BPL households relative to if there was no treatment in the first effect. The effect of PMUY in increasing LPG access to BPL households was not that large. The result is significant at the 1 percent significance level. The standard error is so small because the sample size is very large. Even for the poorest sample, the treatment effect is small: PMUY increases the LPG intake among the poorest households only by 3.8 percentage points. Again, this estimate is significant at a 1 percent significance level. There are only a few observations out of the common support, that is, less than 100. So, the loss of data due to the exclusion of off-support observation is not an issue in this model.
\par The success of the matching exercise lies with the standardised mean bias between treatment and control covariates after matching should be the minimum; less than 10 percent is the widely accepted mean bias limit. As reported in Table \ref{tab:result}, the mean bias after matching is only less than 2 percent, suggesting that the matching is free from any major flaw. Other metrics to examine the quality of the matching exercise are Rubin's B and Rubin's R, which are mentioned earlier are also within the recommended bounds.

\begin{table}[h]
\centering
\begin{threeparttable}
\caption{Impact of PMUY on Firewood Usage}
\label{tab:result2}
\begin{tabularx}{\textwidth}{Xllll}
\toprule
& \multicolumn{2}{c}{Full Sample} & \multicolumn{2}{c}{Poorest Sample} \\
\cmidrule(lr){2-3} \cmidrule(lr){4-5}
& Pre Treatment & Post Treatment & Pre Treatment & Post Treatment \\
\midrule
ATT (Treatment Group)  & 0.618         & 0.459            & 0.823           & 0.703 \\
ATT (Control Group) & 0.589         & 0.452            & 0.791           & 0.720 \\
Difference in ATT & 0.029         & 0.007            & 0.032           & -0.016   \\
S.E & 0.002         & 0.003            & 0.004           & 0.005 \\
t-stat & 12.27         & 2.56             & 8.3             & -3.26 \\
\midrule
Treatment Effect     & \multicolumn{2}{l}{ -0.022 *** } & \multicolumn{2}{l}{-0.048***} \\
Standard Error     & \multicolumn{2}{l}{ 0.003 } & \multicolumn{2}{l}{0.006} \\
\midrule
On support Untreated & 349318      & 194726         & 62463          & 36909 \\
On support Treated    & 218753      & 226180         & 67984          & 68828 \\
\midrule
Mean Bias & 1.6           & 2.1              & 1.2             & 1.2 \\
Rubin's B & 5.4           & 6.6              & 4.5             & 5.7  \\
Rubin's R & 1.01          & 0.9              & 1.08            & 1.07 \\
\bottomrule
\end{tabularx}
\begin{tablenotes}[para,flushleft]
\item \textbf{Note}: The treatment Effect is the difference-in-difference of the ATT from pre and post-treatment period. Sampling variance and standard errors are calculated separately.
\item \textbf{***} $p < 0.01$, \textbf{**} $p < 0.05$,  \textbf{*} $p < 0.1$
\end{tablenotes}
\end{threeparttable}
\end{table}

\par The results from table \ref{tab:result2} suggest a small but significant impact of PMUY on firewood reduction among BPL households in India. PMUY reduced firewood consumption among  BPL households by 2.2 percentage points relative to what would have been for BPL households in the absence of any policy intervention and is significant at a 1 percent significance level. The increase in LPG access due to PMUY is also 2.2 percentage points.  Since the observation on support is large, the standard error is minute, as the estimate's precision increases as the sample size increases. For the poorest sample, the effect is a reduction of firewood consumption by 2.5 percentage points. The mean bias between treated and controlled covariates lies well below the limit, suggesting the quality of the matching exercise. Both Rubin's B and R fall within acceptable limits, indicating that the matching process successfully balanced observable covariates.

\begin{table} [h]
\centering
\begin{threeparttable}
\caption{ATT of PMUY on LPG Access by Zone}
\label{tab:result3}
\begin{tabularx}{\textwidth}{XXXX}
\hline
Zone & ATT & SE & $\beta - \beta^*$ \\
\hline
North         & 0.056***  & 0.010 & 0.035*** \\
North Eastern & -0.026*** & 0.007& -0.047*** \\
Central       & -0.051*** & 0.009  & -0.072***  \\
East          & -0.017 & 0.012    & -0.038*** \\
West          & 0.039***  & 0.010 & 0.018*\\
South         & 0.046***  & 0.014  & 0.025*\\
\hline
\end{tabularx}
\begin{tablenotes}[para,flushleft]
\small
\item \textbf{Note}: ATT is the difference-in-difference of ATTs from pre and post treatment period. t-stat is computed separately using sampling variance. * represents significance level.
\item \textbf{***} $p < 0.01$, \textbf{**} $p < 0.05$,  \textbf{*} $p < 0.1$
\end{tablenotes}
\end{threeparttable}
\end{table}

\par Table \ref{tab:result3} reports the ATT estimates of PMUY on the LPG access in different zones in India. The treatment effects are significant in all regions except for the Eastern zone, which includes the states of Bihar, Jharkhand, Odisha, and West Bengal. These Eastern zone states are predominantly economically disadvantaged with low per capita income. In the North Zone, comprising various states, PMUY has had the most significant impact, increasing LPG uptake by 5.6 percentage points. The South Zone, encompassing states like Andhra Pradesh, Karnataka, Kerala, Tamil Nadu, and Telangana, follows closely with a substantial impact of 4.6 percentage points. These results are highly statistically significant at the 1 percent level. However in the Northeastern and Central Zones, PMUY has led to a reduction in LPG uptake. In the Northeastern zone, there is a decrease of 2.6 percentage points, while the Central zone shows a more significant reduction of 5.1 percentage points. This implies that, on average, PMUY has decreased the likelihood of BPL households gaining access to LPG by 5.1 percentage points in the Central zone and 2.6 percentage points in the North East zone. The Central Zone includes states like Chattisgarh, Madhya Pradesh, Uttarakhand, and Uttar Pradesh. Conversely, in the West Zone, there has been a notable increase of 3.9 percentage points in LPG consumption among BPL households compared to if there had been no treatment. This result is also statistically significant at the 1 percent level. 
\par The reason why the PMUY has a negative causal impact in the North East and East zone is mainly because of the high proportion of ST households. From the estimates reported in table \ref{castetable}, it is evident that the ST households are at a disadvantage in access to LPG. To understand the reason why these regions show a negative effect, a separate sub-sample analysis for the North East, East and Central zones has been carried out.  The result suggests that for the Eastern zone, PMUY has a small but significant impact of 0.6 percentage points on ST households but not a significant impact for the other category households. The minor effect observed in ST households and the lack of any impact on households in other categories collectively render the overall statistically insignificant impact for the entire region. For the Central zone, the impact is negative for all caste categories but significant only for the non-ST households. A further deep analysis into the reasons why every category of households is at a disadvantage in accessing LPG in the Central region. For the Nort-East sample again, the impact is statistically significant only for SC households. For all other categories, the impact is statistically insignificant. The result indicates that the SC household has a negative causal impact of 4.2 percentage points in accessing LPG due to PMUY \footnote{The result for the separate sub-sample analysis is provided in Appendix A}. 

\par A t-test is conducted to determine whether the ATT estimated for different zones is statistically different from the ATT estimated for the whole country. This is reported in the fourth column of the table \ref{regiontable}. The null hypothesis is $\beta = \beta^{*}$. And the alternative hypothesis is it is not. The $\beta$ is the estimated ATT of the different zones and $\beta^{*}$ is the ATT for the whole sample which is reported in the table \ref{tab:result}. However, based on the t-test, the null hypothesis can only be rejected for the North, North Eastern, Central and Eastern samples at a 1 percent significance level.  This means that the ATT estimates are significantly different from that of the ATT estimate of the whole sample.

\begin{table}[ht]
  \centering
  \begin{threeparttable}
    \caption{Region-wise ATT of PMUY on LPG access}
    \label{regiontable}
    \begin{tabularx}{\textwidth}{XXXX}
    \toprule
    Region & ATT & SE & $\beta - \beta^*$ \\
    \midrule
    Urban & 0.003 & 0.006 & -0.018***\\
    Rural & 0.024*** & 0.004 & 0.003 \\
    \bottomrule
    \end{tabularx}%
    \begin{tablenotes}[flsuhleft, para]
      \small
      \item Note : \textbf{***} $p < 0.01$, \textbf{**} $p < 0.05$,  \textbf{*} $p < 0.1$, 
    \end{tablenotes}
  \end{threeparttable}
\end{table}

\par Table \ref{regiontable} shows that the impact of PMUY is larger in the rural rather than in the urban areas because 81 percent of the total BPL households in the sample are in rural areas. On average the impact of PMUY on LPG access among BPL households in rural areas increased by 2.4 percentage points. But for the urban sample, PMUY do not have a significant Average Treatment Effect. This may be because of the low representation of BPL households in urban areas. Out of the 19 percent of BPL households living in urban areas, only 5.93 percent of households do not have access to LPG in urban areas. This might be why the PMUY has no significant impact in the urban areas. The t-test to determine whether the urban and rural estimates are statistically different from the national estimate is also provided. The urban estimate is statistically different from the national estimate, but the rural estimate is not statistically different from the national estimate, because most of the treatment groups are in rural areas. 

\begin{table}[htbp]
  \centering
  \begin{threeparttable}
    \caption{Religion-wise ATT of PMUY on LPG access}
    \label{religiontable}
    \begin{tabularx}{\textwidth}{XXXX}
    \toprule
    Religion & ATT & SE & $\beta - \beta^*$  \\
    \midrule
    Hindu     & 0.030***  & 0.004 & 0.009**
 \\
    Muslim    & 0.057***  & 0.011 & 0.036***
   \\
    Christian & -0.040*** & 0.010&  -0.061***
  \\
    Sikh      & 0.029  & 1.130   &   0.008
 \\
    \bottomrule
    \end{tabularx}
   \begin{tablenotes}[flsuhleft, para]
      \small
      \item Note : \textbf{***} $p < 0.01$, \textbf{**} $p < 0.05$,  \textbf{*} $p < 0.1$, 
    \end{tablenotes}
  \end{threeparttable}
\end{table}

\par The table \ref{religiontable} gives the ATT of the PMUY on major religions in India. The treatment effect has more impact on the Muslim group. The ATT of the Muslim group suggest that, on average, the causal impact of PMUY on LPG access among Muslim households has increased by 5.7 percentage points relative to that if there was no treatment. Hindu households have an impact of  3 percentage points relative to what if there was no treatment. The estimate for the Christian group is particularly interesting because it displays a negative sign for the estimated treatment coefficient. This result should be read in the backdrop of the results from table \ref{tab:result3} that the impact of the treatment in the Northeast region is negative. And around 17 percent of the households in the North East zone are Christian. Also, almost 85 percent of Christian households belong to the scheduled tribe in the sample. When the PSM is estimated for Christian religion without ST households and only ST households separately, the impact of PMUY is positive for non-ST households, while it is negative for ST households in the Northeast Region (Appendix A). This might be one reason why the Christian households showing a negative impact on LPG access by PMUY in this sample. The LPG access of Christian households that belong to the scheduled tribe has increased from 31 percent in 2014-15  to 39 percent in 2019-20, while the households which do not belong to the scheduled tribe household increased their LPG share from 55 percent in 2014-15 to 74 percent in 2019-20. The ATT for both years are negative for ST households, while it is positive for the non-ST households (Appendix A).

\par Sikh households show no causal impact by PMUY. Again, for the Sikh community, the population in the sample is divided in such a way that almost equal proportion of people belong to the Scheduled caste and Upper Caste. When the causal impact of Sikh households is estimated separately for the households that belong to the SC and Non-SC categories, the treatment effects are insignificant for both SC and non-SC households. So, the PMUY has no causal impact on any category of Sikh households.

The difference from the whole sample estimate is statistically significant for all religions except the Sikh religion.

\begin{table}[htbp]
  \centering
  \begin{threeparttable}
    \caption{Caste-wise ATT of PMUY on LPG access}
    \label{castetable}
    \begin{tabularx}{\textwidth}{XXXX}
    \toprule
    Caste & ATT & SE & $\beta - \beta^*$ \\
    \midrule
    SC & 0.041*** & 0.007& 0.020***\\
    ST & -0.007 & 0.007 & -0.028*** \\
    OBC & 0.033*** & 0.008 & 0.020*** \\
    None of the above & 0.031*** & 0.008 & 0.012 \\
    \bottomrule
    \end{tabularx}%
    \begin{tablenotes}[flsuhleft, para]
      \small
      \item Note : \textbf{***} $p < 0.01$,\textbf{**} $p < 0.05$ , \textbf{*} $p < 0.1$, 
    \end{tablenotes}
  \end{threeparttable}
\end{table}

\par Table \ref{castetable} shows that PMUY has a maximum causal impact among the OBC households of 4.1 percentage points, followed by SC households. The ATT estimate of the ST households is not significant. This is mainly because the ST households have a significant proportion of the households in the North Eastern and Eastern Zone. The causal impact of the PMUY is negative in the North Eastern Region. In a subsample analysis for the North Eastern sample without ST households, the ATT estimate is positive and in a sub-sample analysis with only ST households in the North Eastern region, the ATT estimate was negative and insignificant. Similarly, In a similar matching exercise for the Eastern sample, the estimate shows that for ST households, the ATT estimate is negative and significant. However, the result again becomes insignificant for non-ST households.  Either way, both Eastern and North Eastern regions are disadvantaged in accessing LPG; among the households, the ST households are disadvantaged more than any other category households. 

\begin{table} [h]
\centering
\begin{threeparttable}
\caption{ATET of PMUY using IPW and AIPW estimator}
\label{tab:result4}
\begin{tabularx}{\textwidth}{XXX}
\hline
            & LPG Access & Firewood Consumption \\
\hline
ATET (IPW)  & 0.008***   & -0.013***             \\
            & (0.002)   & (0.002)            \\
ATET (AIPW) & 0.008***    & -0.013***             \\
            & (0.002)   & (0.002)             \\
\hline
\end{tabularx}
\begin{tablenotes}[para,flushleft]
\small
\item  \textbf{Note}: Robust standard errors in parenthesis. \textbf{***} $p < 0.01$, \textbf{**} $p < 0.05$, \textbf{*} $p < 0.1$
\end{tablenotes}
\end{threeparttable}
\end{table}

\par As a robustness check, the Average Treatment Effect of PMUY was estimated using the Inverse Probability weight (IPW) and the Augmented Inverse Probability Weight (AIPW) Estimator. The results are reported in table \ref{tab:result4}. Using the IPW estimator, the estimated ATET should be equal to ATT after successful randomisation. The estimates are approximately equal for both  ATET of IPW and AIPW. This suggests that the ATE estimates are consistent. Since the AIPW is the most efficient of all estimators because of its doubly robust properties \citep{sloczynski2018general}, the estimated ATE can be treated as an unbiased estimate. The estimated ATE can be interpreted as the causal impact of PMUY on the treatment group has led to an average increase of LPG access by 0.8 percentage points relative to what if there was no treatment in the first place among the treated BPL households. And for firewood usage, the average effect is even more negligible at only  0.05 percentage points. Both effects are statistically significant at a 1 percent level. Since the AIPW estimator is doubly robust, if either the outcome model or the model to estimate the probability of the treatment is correctly specified, then the Average Treatment on the Treated is efficient. 

\par Since the ATT estimate from the PSM model and ATET from IPW and AIPW predict the treatment effects in the same direction and the magnitude is similar, it should be concluded that the impact of PMUY is indeed small among BPL households.

\begin{table}[h]
\centering
\begin{threeparttable}
\caption{Rosenbaum Bound Test Sensitivity Analysis using MH-stat}
\label{sensitivitytable}
\begin{tabularx}{\textwidth}{XXXXXXXXX}
\toprule
& \multicolumn{4}{c}{Pre Treatment} & \multicolumn{4}{c}{Post Treatment} \\
\cmidrule(lr){2-5}\cmidrule(lr){6-9}
$\Gamma$ & Q\_mh+ & Q\_mh- & p\_mh+ & p\_mh- & Q\_mh+ & Q\_mh- & p\_mh+ & p\_mh- \\
\midrule
1.0   & 58.827 & 58.827 & 0.000 & 0.000 & 35.866 & 35.866 & 0.000 & 0.000 \\
1.1 & 69.686 & 48.042 & 0.000 & 0.000 & 46.788 & 24.983 & 0.000 & 0.000 \\
1.2 & 79.681 & 38.249 & 0.000 & 0.000 & 56.803 & 15.072  & 0.000 & 0.000 \\
1.3 & 88.954 & 29.274 & 0.000 & 0.000& 66.063 & 5.967 & 0.000 & 0.000 \\
1.4 & 97.618 & 20.986 & 0.000 & 0.000 & 74.682 & 2.450 & 0.000 & 0.007 \\
1.5 & 105.761 & 13.283 & 0.000 & 0.000 & 82.753  & 10.296  & 0.000 & 0.000 \\
1.6 & 113.451 & 6.085 & 0.000 & 0.000 & 90.348 & 17.640 & 0.000 & 0.000 \\
1.7 & 120.747 & 0.665 & 0.000 & 0.253    & 97.526 & 24.547 & 0.000 & 0.000 \\
1.8 & 127.694 & 7.038  & 0.000 & 0.000 & 104.336 & 31.069 & 0.000 & 0.000\\
1.9 & 134.331 & 13.069   & 0.000 & 0.000        & 110.819 & 37.249 & 0.000 & 0.000 \\
2.0   & 140.692 & 18.794  & 0.000& 0.000        & 117.009 & 43.126 & 0.000 & 0.000 \\
\bottomrule
\end{tabularx}
\begin{tablenotes}[para,flushleft]
\item Note: $\Gamma$: odds of differential assignment due to unobserved factors.
\par mh: is the MH stat for overestimation(+) and underestimation(-)
\par p\_mh+ and p\_mh- are the p-values 
\end{tablenotes}
\end{threeparttable}
\end{table}

\par The results from Table \ref{sensitivitytable} indicate that the potential impact of unobservables is minimal in this model. The $\Gamma$ represents the odds of differential assignment due to unobserved variables. When $\Gamma$ is 1, the model has complete randomness. It suggests that if two random households are selected, the probability of getting the treatment is equal, as no unobservable factors affect their probability of getting treatment. As the $\Gamma$ increases, the potential unobservable differences between the households increase; therefore, the overestimation and underestimation of the treatment effect also increases. Since ATE is positive, the potential bias can most possibly overestimate the ATE.  But what is indicated from table \ref{sensitivitytable} is that even if there are potential unobservable differences between the households, the odds that it will overestimate or underestimate the treatment effects are insignificant. In other words, the null hypotheses of no treatment effects are significantly rejected in this model both in pre and post-treatment matching. The Rosenbaum bound test only indicates what happens if there is any unobservable selection bias in the model. It does not say whether there is any unobservable selection bias in the model or the exact magnitude of the bias. However, it can be ensured that the unobservable will not affect the treatment effects in this model even if the potential unobservable factors affect the odds of treatment assignment even up to a factor of 2. Even when the $\Gamma$ is 2, when the odds of the differential assignment are doubled, the null hypothesis of no treatment effect is rejected. This sensitivity analysis concludes that the treatment effect estimated using the observable cofounders is insensitive to the unobservable confounders, and there is no problem of unobservable selection bias.

\subsection{Limitations of the Study}
\par One of the primary limitations of this study stems from the absence of a policy (PMUY) variable within the National Family Health Survey (NFHS) dataset. Consequently, the study relies on an intent-to-treat methodology, which introduces a scenario where BPL households that already possess an LPG connection are also counted as treated, potentially biasing the estimates. However, it's worth noting that LPG utilisation among households categorised as Below Poverty Line (BPL) is quite minimal, mitigating this bias concern. Even if the treatment assignment is imperfect due to the intent-to-treat analysis, the results can still be interpreted as the LPG utilisation among BPL households between 2014-15 and 2019-20 in the absence of treatment. Nevertheless, the outcome remains concerning, as there was only a small growth in LPG adoption among BPL households. This highlights that regardless of the treatment group's imperfections, BPL households face challenges in accessing LPG.

\par Another significant challenge encountered in the analysis pertains to the definition of the cooking fuel variable itself. In the NFHS questionnaire, households are asked about their primary source of cooking fuel. If households have access to LPG but do not primarily use it for cooking, NFHS categorises them as not having LPG access. This poses an issue when attempting to analyse LPG access. However, it's important to emphasise that the success of PMUY is ultimately measured by the extent to which people use LPG as their primary cooking fuel. Consequently, the variable definition aligns well with the study's focus on LPG access. Another potential drawback of the model is the lack of a specific household expenditure variable, with the wealth index serving as a substitute in its place.

\section{Conclusion and Policy Implications}
\par The Pradhan Mantri Ujjwala Yojana demonstrably improved LPG accessibility among BPL households in India, achieving a modest but statistically significant 2.1 percentage point increase in LPG consumption alongside a corresponding decrease in firewood dependence. This shift in energy consumption patterns represents a crucial step towards cleaner cooking fuel adoption and improved household health outcomes. However, the findings also reveal regional disparities in program effectiveness, with the North, West, and South experiencing greater gains compared to the East and North East zones. Moreover, the study underscores the need for more ambitious and sustained policy interventions to ensure a wider and long-lasting impact.

\par The analysis of two waves of the NFHS dataset reveals regional disparities in accessing LPG among BPL households, warranting the proposal of tailored regional interventions to address this imbalance. The government's claim of 95 million beneficiaries for the program, accompanied by the results from our analysis, clearly indicates the need for more income support to facilitate an increased and permanent adoption of LPG among poor households. 

\par Looking forward, there are ample opportunities to extend this research. We did not consider the time use associated with the different fuels. An intriguing avenue involves exploring the impact of PMUY on respiratory health, particularly among women and children who often bear the adverse effects of harmful smoke from traditional cooking methods. Given PMUY's explicit objective of reducing respiratory ailments resulting from such exposures, delving into this area could yield valuable insights into the program's comprehensive health benefits. Improved respiratory health could also lead to better outcomes for education and a greater earning capacity. All this is postponed to future research.

\clearpage
\bibliographystyle{apalike}
\bibliography{reference}
\clearpage

\appendix
\section{Additional results}

\begin{table}[h]
\caption{Summary Statistics: Extended}
\begin{tabularx}{\linewidth}{lXll}
\toprule
\textbf{Variables} & \textbf{Variable Description} & \textbf{Min} & \textbf{Max} \\
\hline
 Age & Age of the Household Head & 10 & 98 \\
HHsize & Number of members in the households & 1 & 41 \\
Urban & 1. Urban, 2. Rural & 1 & 2 \\
Gender & Gender of the household head. 1-Male, 2-Female & 1 & 2 \\
Zone & 1. North Zone, 2. North East Zone, 3. Central Zone, 4. East Zone, 5. West Zone, 6. South Zone & 1 & 6 \\
Cooking fuel & 1. Electricity, 2. LPG, 3. Biogas, 5. Kerosene, 6. Coal, 7. Charcoal, 8. Wood, 9. Straw, 10. Agricultural crop, 11. Animal dung, 95. No cooking in the Household, 96. Others & 1 & 96 \\
Wealth Index & 1. Poorest, 2. Poorer, 3. Middle, 4. Richer, 5. Richest & 1 & 5 \\
Religion & 1. Hindu, 2. Muslim, 3. Christian, 4. Sikh, 5. Buddhist, 6. Jain, 7. Jewish, 8. Parsi, 9. No religion, 96. Other & 1 & 96 \\
Caste & 1. SC, 2. ST, 3. OBC, 4. None of the above, 8. Don't Know & 1 & 8 \\
LPG Access & 0. No, 1. Yes & 0 & 1 \\
\bottomrule
\end{tabularx}
\end{table}

\begin{table}[h]
\caption{Probit Results for matching: pre-treatment}
\begin{tabularx}{\textwidth}{XXXX}
\toprule
\textbf{Variable} & \textbf{Coefficient} & \textbf{SE} & \textbf{p-value} \\
\midrule
State & -0.029 & 0.000 & 0.000 \\
Age & 0.003 & 0.000 & 0.000 \\
Religion & -0.002 & 0.000 & 0.000 \\
Caste & -0.104 & 0.002 & 0.000 \\
Educ & -0.127 & 0.002 & 0.000 \\
Wealth index & -0.181 & 0.002 & 0.000 \\
Urban & 0.172 & 0.004 & 0.000 \\
Gender & 0.011 & 0.005 & 0.029 \\
Hhsize & 0.023 & 0.001 & 0.000 \\
Constant & 0.670 & 0.015 & 0.000 \\
\bottomrule
\end{tabularx}
\caption*{\raggedright Note: The probit coefficient for treatment assignment pre-treatment for the estimation of results in the table \ref{tab:result2} }
\end{table}

\begin{table} [h]
\caption{Probit Results for Matching: Post-treatment}
\begin{tabularx}{\textwidth}{XXXX}
\toprule
\textbf{Variable} & \textbf{Coefficient} & \textbf{SE} & \textbf{p-value} \\
\midrule
State & 0.027 & 0.000 & 0.000 \\
Age & -0.001 & 0.000 & 0.000 \\
Religion & 0.003 & 0.000 & 0.000 \\
Caste & -0.112 & 0.002 & 0.000 \\
Educ & -0.207 & 0.002 & 0.000 \\
Wealth index & -0.097 & 0.002 & 0.000 \\
Urban & 0.397 & 0.005 & 0.000 \\
Gender & -0.056 & 0.005 & 0.000 \\
Hhsize & 0.037 & 0.001 & 0.000 \\
Constant & -0.349 & 0.018 & 0.000 \\
\bottomrule
\end{tabularx}
\caption*{\raggedright Note: The probit coefficient for treatment assignment post-treatment for the estimation of results in the table \ref{tab:result2} }
\end{table}

\begin{table}[h]
\begin{threeparttable}
\caption{ The ATT Estimates for different Samples of Christian Religion}
\begin{tabularx}{\textwidth}{XXX}
\hline
\textbf{Household category} & \textbf{ATT} & \textbf{SE} \\
\midrule
For ST Household& -0.045*** & 0.011  \\
For Non-ST Household & 0.047** & 0.024
\\
\hline
\end{tabularx}
\begin{tablenotes}
\item Note: \textbf{***} $p < 0.01$, \textbf{**} $p < 0.05$, \textbf{*} $p < 0.1$.
\end{tablenotes}
\end{threeparttable}
\end{table}

\begin{table}[h]
\begin{threeparttable}
\caption{The ATT Estimates for different Samples of Sikh Religion}
\begin{tabularx}{\linewidth}{XXX}
\hline
\textbf{Household category} & \textbf{ATT} & \textbf{SE} \\
\hline
For SC Household & -0.015
 & 0.033
 \\
For non-SC Household & -0.038
 & 0.039 \\
\hline
\end{tabularx}
\begin{tablenotes}
\item Note: \textbf{***} $p < 0.01$, \textbf{**} $p < 0.05$, \textbf{*} $p < 0.1$.
\end{tablenotes}
\end{threeparttable}
\end{table}

\begin{table}[h]
\caption{The ATT Estimates for Different Eastern Zone Samples}
\begin{tabularx}{\textwidth}{XXX}
\toprule
\textbf{Sample} & \textbf{ATT} & \textbf{SE} \\
\midrule
Only With ST Households   & 0.006***  & 0.020 \\
Without ST Households     & -0.017 & 0.016 \\ \bottomrule
\end{tabularx}
\caption*{ \raggedright Note: \textbf{***} $p < 0.01$, \textbf{**} $p < 0.05$, \textbf{*} $p < 0.1$}
\end{table}

\begin{table}[h]
\caption{The ATT Estimates for Different Central Zone Samples}
\begin{tabularx}{\textwidth}{XXX}
\toprule
\textbf{Sample} & \textbf{ATT} & \textbf{SE} \\
\midrule
Only With ST Households   & -0.017  & 0.031 \\
Without ST Households     & -0.020** & -2.058 \\ \bottomrule
\end{tabularx}
\caption*{ \raggedright Note: \textbf{***} $p < 0.01$, \textbf{**} $p < 0.05$, \textbf{*} $p < 0.1$}
\end{table}

\begin{table}[h]
\caption{The ATT Estimates for Different North Eastern Zone Samples}
\begin{tabularx}{\textwidth}{XXX}
\toprule
\textbf{Sample} & \textbf{ATT} & \textbf{SE} \\
\midrule
Only With ST Households   & 0.003  & 0.011 \\
Without ST Households     & -0.028*** & 0.010 \\
Only With SC Households   & -0.042** & 0.019 \\
Only With OBC Households  & -0.020 & 0.015 \\ \bottomrule
\end{tabularx}
\caption*{ \raggedright Note: \textbf{***} $p < 0.01$, \textbf{**} $p < 0.05$, \textbf{*} $p < 0.1$}
\end{table}

\end{document}